# Structural, electronic, and magnetic properties of tris(8-hydroxyquinoline)iron(III) molecules and its magnetic coupling with ferromagnetic surface: first-principles study

Wei Jiang,[1] Miao Zhou,[1] Zheng Liu,[1] Dali Sun,[2] Z. Valy Vardeny[2] and Feng Liu[1*]

[1]Department of Materials Science & Engineering, University of Utah, Salt Lake City, UT 84112.
[2]Department of Physics, University of Utah, Salt Lake City, UT 84112.

**ABSTRACT:** Using first-principles calculations, we have systematically investigated the structural, electronic, and magnetic properties of facial (*fac-*) and meridional (*mer-*) tris(8-hydroxyquinoline)iron(III) (Feq$_3$) molecules and their interaction with ferromagnetic substrate. Our calculation results show that for the isolated Feq$_3$, *mer*-Feq$_3$ is more stable than the *fac*-Feq$_3$; and both Feq$_3$ isomers have a high spin-state of 5 $\mu_B$ as the ground state when an on-site Hubbard-U term is included to treat the highly localized Fe 3$d$ electrons; while the standard DFT calculations produce a low spin-state of 1 $\mu_B$. These magnetic behaviors can be understood by the octahedral ligand field splitting theory. Furthermore, we found that *fac*-Feq$_3$ has a stronger bonding to the Co surface than *mer*-Feq$_3$ and an anti-ferromagnetic coupling was discovered between Fe and Co substrate, originating from the superexchange coupling between Fe and Co mediated by the interface oxygen and nitrogen atoms. These findings suggest that Feq$_3$ molecular films may serve as a promising spin-filter material in spintronic devices.

PACS number(s):71.20.Rv, 71.70.Ch, 75.30.Et, 77.84.Jd

# I. INTRODUCTION

Trischelate oxyquinoline octahedral metal complexes (Mq$_3$) have been widely studied for their potential applications in organic solar cells, light emission diodes (OLEDs), and data storage and communication devices [1,2]. One typical example, Alq$_3$ has been extensively studied for OLEDs transport as well as for spin-valve materials in organic systems [3,4]. Another molecule Feq$_3$ [5-7], with the same structure as Alq$_3$, has a paramagnetic nature that is in sharp contrast to Alq$_3$ (nonmagnetic). In recent experiments [8], Feq$_3$ was shown to greatly enhance the performance of organic spintronic devices. However, to date, many basic properties of this molecule have yet to be studied. One early theoretical work [5] had studied some magnetic properties of Feq$_3$ molecules, however, the localized electrons of Fe in Feq$_3$, which have a dramatic influence on the calculation results, were not properly treated. On the other hand, it has been shown that the hybrid ferromagnetic metal (FM)/organic interfaces have a tremendous influence on the electrical transport and magnetic properties of the organic spintronic devices [3,4,7-10]. Unlike nonmagnetic Mq$_3$/FM systems, such as Alq$_3$, Gaq$_3$, and Inq$_3$ that have been broadly studied both theoretically and experimentally [11,12], studies of magnetic Mq$_3$/FM systems have been very limited. This has motivated us to perform a systematic study of the structural, electronic, and magnetic properties of facial and meridional isomers of Feq$_3$, as well as their interaction with a ferromagnetic Co substrate, using first-principles computational methods based on density functional theory (DFT). Combining the orbital hybridization theory [13], crystal field theory, and the DFT results, we have comprehensively explained their different magnetic behaviors in different situations. The mechanisms of the magnetic coupling and interactions between Feq$_3$ and Co substrate are carefully examined.

## II. COMPUTATIONAL DETAILS

We have carried out first-principles calculations based on DFT. As implemented in Vienna *ab initio* simulation package (VASP) code, the projector augmented wave and the generalized gradient approximation [14,15] exchange-correlation potentials method were used. In order to better describe the system with the localized (strong correlated) 3$d$ electrons of Fe, an additional on-site Hubbard-U correction term [16,17] was added with the different U values for Fe (U=1-6 eV, J=0.9 eV) [18,19] tested. The plane wave cutoff energy was set to 400 eV and the valence configuration $3d^7 4s^1$ was used for Fe atom. A large $20 \times 20 \times 20$ Å$^3$ supercell was used to simulate isolated Feq$_3$ molecules to minimize interactions between neighboring molecules. A four-layered cobalt slab with $6 \times 6$ cobalt atoms in each layer was used to simulate the cobalt hcp(0001) surface with bottom two layers of atoms fixed during the structural relaxation. A 15 Å vacuum layer along the $z$ direction was included to eliminate the interactions between slabs. For adsorption of Feq$_3$ on the cobalt surface (Feq$_3$/Co), the DFT-D2 method [20] was applied to describe the van der Waals interactions, which have been shown to have a dramatic influence on the molecular adsorption energies and geometries [21,22]. For the brillouin zone sampling, the Γ point was used for the isolated Feq$_3$ molecule, and a $3 \times 3 \times 1$ Monkhorst-Pack [23] $k$ mesh was used for the cobalt and Feq$_3$/Co systems. The structural relaxations were carried out until the force on each atom was less than 0.03 eV/ Å.

## III. RESULTS AND DISCUSSIONS

Same as the structure of Alq$_3$, one central iron atom surrounded by three 8-hydroxyquinoline ligands makes up one Feq$_3$ molecule, which has two types of isomers, facial and meridional [see Fig. 1(a,b)]. For the isolated Feq$_3$ molecule, structural relaxation was performed to obtain the ground state geometry. The center Fe atom is in a distorted $FeN_3O_3$ octahedron with oxygen atoms (nitrogen atoms) forming right-angled triangles and regular triangles for *mer*-Feq$_3$ and *fac*-Feq$_3$, respectively. As shown in Table I, our calculation results agree very well with previous experimental results [6,7]. The calculated bond lengths of Fe-O and Fe-N for *mer*-Feq$_3$ varies in the range of 1.960-1.994 and 2.158-2.217 Å, respectively; while for the *fac*-Feq$_3$, the bond lengths of Fe-O and Fe-N are all the same of 1.956 and 2.229 Å, respectively, implying a higher structural symmetry than *mer*-Feq$_3$. In order to compare the stability of the two Feq$_3$ isomers, we carried out the calculations with and without on-site U term. Our results of both methods gave the total energy of *mer*-Feq$_3$ to be about 0.13 eV lower than that of *fac*-Feq$_3$, indicating that *mer*-Feq$_3$ is thermodynamically more stable in the isolated state.

For the magnetic properties of two isomers, the value of on-site U term plays a very important role in determining the ground spin-state. We calculated the energies as a function of magnetic moment with and without the on-site U term by fixing the magnetic moment at different values for comparison. From table II, it is evidently seen that before adding the on-site U term on Fe, the ground spin-state is the low spin-state (1 $\mu_B$) as shown by a previous calculation [5]. However, after applying the on-site U term, the ground state switches to high spin-state (5 $\mu_B$) that is consistent with the early experimental results [7]. Moreover, we calculated the magnetic moment as a function of different U values. From Fig. 1(c), it is clear that both Feq$_3$ isomers have two spin-states (5 $\mu_B$, high spin-state; 1 $\mu_B$, low spin-state), and the ground spin-state changes from low spin-state to high spin-state as the value of the on-site U term increases. Furthermore, when the U value is greater than 6 eV, two Feq$_3$ isomers have identical magnetic behavior. Our calculations show that the total magnetic moment of Feq$_3$ does not change at higher U values, only the magnetic moment of Fe atom increase continuously. For

*fac*-Feq$_3$, the ground spin-state keeps in a high spin-state when the U value is larger than 1 eV. However, some oscillation between high and low spin-state appeared for *mer*-Feq$_3$ as the U value increases. For the calculation of the Co/Feq$_3$ systems, a 6 eV U term was used to keep the high spin-state and to make the calculation results of two Feq$_3$ isomers comparable.

To further understand the origin of the observed magnetic behaviors, we calculated and analysed the density of states (DOS) for different magnetic spin-states as shown in Fig. 2. Because the shapes of the DOS are almost the same for two isomers, only the DOS of the *mer*-Feq$_3$ are discussed here. From Fig. 2(a), it is clear that an exchange splitting exists between the spin-up and spin-down states mainly on Fe atom, which causes an unequal filling of the electrons and gives Feq$_3$ molecule a magnetic property. Comparing the upper and the lower panel of Fig. 2(a), we can see that the exchange splitting of Fe atom for the high spin-state (about 10 eV) is much larger than the low spin-state (less than 1 eV), which leads to different magnetic moments for these two spin-states. Also, the magnetic moment is predominantly localized on the Fe 3*d* orbitals (4.25 $\mu_B$/5 $\mu_B$ or 0.85 $\mu_B$/1 $\mu_B$) with the rest distributed on the neighboring O, N, and C atoms. Our spin distributions [as shown in Fig. 2(b,c)] further confirm these results.

The mechanisms underlying these different magnetic behaviors can be traced back to the molecular ligand splitting associated with the exchange splitting. We analyzed the partial density of states (PDOS) and focusing on the Fe 3*d* orbitals. A schematic view of the energy levels for the *d* orbitals for high (low) spin-state is provided in Fig. 2(d) upper panel (lower panel) to demonstrate the magnetic properties. First, the center Fe atom has three nearest oxygen atoms that need three electrons to close *sp*-shells and three nitrogen atoms that already have the closed *sp*-shells, which requires the Fe atom to transfer three electrons to the neighboring oxygen atoms. Then, the exchange coupling splits spin-up and spin-down states into two parts, and the five degenerated *d* orbitals (xy, xz, yz, x$^2$-y$^2$, and z$^2$) split into $e_g$ (x$^2$-y$^2$ and z$^2$) and $t_{2g}$ (xy, xz, and yz) two groups due to the crystal splitting. Consequently, three of the twelve C=O and C=N orbitals hybridized with Fe $t_{2g}$ orbitals into $t_{2g}$ bonding and $t_{2g}^*$ anti-bonding state due to the same π symmetry, in which both states have three orbitals, and the $t_{2g}^*$ anti-bonding state has a higher energy than the $e_g$ state. Therefore, three spin-down electrons from the neighboring atoms (N, O, and C) fill the relatively lower $t_{2g}$ state, and three spin-up electrons fill the $t_{2g}^*$ state. For the high spin-state, the larger exchange splitting energy between spin-up and spin-down states makes the five *d* electrons of Fe atom to fill the relatively lower energy band of $t_{2g}$ and $e_g$ bands of the spin-up state, resulting in a total of 5 $\mu_B$ magnetic moment [Fig. 2(d), upper panel]. For the low spin-state of 1 $\mu_B$, it can only happen when the exchange splitting is low, and the spin-up and spin-down states have a small energy difference [Fig. 2(d), lower panel].

    To further study possible applications of this molecule, we carried out the calculations of the adsorption of Feq$_3$ molecules on the cobalt surface. Different absorption configurations of both Feq$_3$ isomers were studied, and particularly the magnetic interaction of the most stable Co/Feq$_3$ configuration was investigated in detail. After the structural relaxation of the cobalt slab, the top surface and the subsurface layer are lowered by 0.15 Å and 0.04 Å, respectively. The average magnetic moment of cobalt is about 1.69 $\mu_B$/atom with the top and the bottom surface layer having 1.73 and 1.77 $\mu_B$/atom, respectively, consistent with the previous calculation results [12]. Then, we considered various adsorption configurations and focused on four geometries that have highest adsorption symmetry with largest adsorption energies which are shown in Fig. 3. To better describe these configurations, the configuration that has more oxygen atoms (nitrogen atoms) bonding with the cobalt surface is denoted as -O (-N), e.g., *fac*-O denotes the configuration that *fac*-Feq$_3$ bonds with the cobalt surface through three oxygen atoms and

*mer*-N denotes the configuration that *mer*-Feq3 bonds with the cobalt surface through two nitrogen atoms. The bonding energy $E_b$ is defined as

$$E_b = E_{mol} + E_{surf} - E_{mol+surf} \qquad (1)$$

where $E_{mol}$, $E_{surf}$, and $E_{mol+surf}$ are the total energy of an isolated Feq3 molecule, the four-layered cobalt slab, and the Co/Feq3 system after structural relaxation, respectively.

The magnetic moment, binding energy, and the energy difference between the four configurations are listed in Table III. It can be seen that all these configurations have the same magnetic moment for Feq3 molecules, and the total magnetic moment varies due to different interaction strength caused by different structural interfaces. The most stable configuration is the *fac*-O that has the highest binding energy of 7.65 eV and total magnetic moment of 241.19 $\mu_B$. Also, the configuration with more oxygen atoms facing the cobalt surface has relatively lower total energy and higher bonding energy for both *fac*- and *mer*-Feq3 isomers, e.g., *fac*-O is more stable than *fac*-N by 0.39 eV, indicating a stronger bonding of Co-O than Co-N, consistent with the experimentally observed results in Alq3 systems [12]. It has been shown above that for isolated Feq3 isomers, *mer*-Feq3 is more stable than *fac*-Feq3. However, surprisingly, when being adsorbed onto the cobalt surface, the Co/*fac*-Feq3 system becomes more stable than the Co/*mer*-Feq3 system. Taken *fac*-N as an example: though it has no oxygen atoms bonding with cobalt surface, it is still more stable than *mer*-N, which can possibly be explained by its high structural symmetry. There might be two main reasons contributing to the stability of the Co/Feq3 system: one is that the hybridization between the oxygen atom and the cobalt surface is preferable; and the second is that the high structural symmetry of the *fac*-Feq3 helps to build a strong bonded interface that decreases the total energy.

The magnetic coupling between this magnetic molecule and the ferromagnetic cobalt surface is crucial for its applications in organic spintronic devices. Therefore, we studied both the ferromagnetic (FM) and antiferromagnetic (AFM) alignments between the Fe and Co spins of the Co/Feq$_3$ system. In order to study the influence of both N and O to the interface state, besides the most stable *fac*-Feq$_3$/Co system, *mer*-Feq$_3$/Co that has two O atoms and one N atom contacting with the Co surface was also studied for comparison (as shown in Fig. 4, similar results are also obtained for the *fac*-Feq$_3$/Co system). Our calculation results show that the AFM is more stable (47 meV for *mer*-Feq$_3$/Co; 71 meV for *fac*-Feq$_3$/Co) than the FM configuration for both systems, indicating a preferred AFM coupling between the Fe and Co. To better understand this observation, we calculated the DOS of both FM and AFM configurations and analyzed the PDOS for the Fe, Co, N, and O atoms [see Fig. 4(a)]. There is little direct 3$d$-3$d$ overlap between the Fe and Co states, but the *p-d* (Fe-N, Fe-O and Co-N, Co-O) hybridizations are clearly seen, according to the overlap of their energy states. Moreover, the states of oxygen are more widely broadened than the states of nitrogen below the Fermi level, which confirms the preference for the Co-O bonds. Because of these strong hybridizations, a clear increase between spin-up and spin-down splitting can be seen for the oxygen states in AFM configuration than in the FM configuration; and the direction of the spin-polarization of O follows with the cobalt surface instead of the Fe. Hence the oxygen and nitrogen atoms act as bridges that transmit an indirect interaction between Fe and Co. By further analyzing the magnetic distribution of the interface atoms [see Fig. 4(b)], the magnetic moments of O atoms that bond with the cobalt surface were found to decrease and change signs due to the charge transfer from the substrate, and the magnetic moment of Fe decreases accordingly. However, the magnetic moment of the cobalt atoms that bond with O and N atoms increases dramatically compared to other cobalt atoms. These findings further confirm an indirect interaction between Fe and Co through O and N atoms.

Thus, we identified that the AFM coupling between Fe and Co is caused by a kind of superexchange coupling (Fe-O,N-Co) mediated by the interfacing oxygen and nitrogen atoms that bond strongly with both Fe and Co. Interestingly, this is different from the previously found Fe(OEP)/Co system, which showed similar superexchange coupling, but having an FM coupling between Fe and Co. The mechanisms underlying this difference can be understood by the Goodenough-Kanamori rules [24]. For the superexchange couplings that have M-O-M bonds, the sign of the exchange coupling is determined by the orbital degeneracy and occupancy of the $3d$ states. Instead of the 90º Fe-O-Co angle bond in Fe(OEP)Co system, we find a nearly 120º Fe-O,N-Co angle bond in our $Feq_3$/Co system, which gave an AFM coupling state between Fe and Co. Therefore, together with the FM coupled Fe(OEP)/Co system [25], our newly found AFM coupled $Feq_3$/Co system may provide a new class of spin-dependent molecular electronics.

## IV. CONCLUSIONS

In this paper, the structural, electric, and magnetic properties of $Feq_3$ were systematically studied through first-principles calculations. The ground state was found to be the high spin-state (5 $\mu_B$) after applying the on-site U term on Fe element, rather than the low spin-state (1 $\mu_B$) reported before [5]. The physics of crystal splitting and exchange coupling for the magnetic behaviors were explained in detail. In addition, due to the high structural symmetry and the preference for the Co-O bonds over the Co-N bonds, the thermodynamically unfavorable isolated *fac*-$Feq_3$ isomer was found to have a stronger bonding to the cobalt substrate than the *mer*-$Feq_3$. Furthermore, an AFM coupling between Fe and Co was found in our system rather than the FM coupling in Fe(OEP)/Co system. Noticeably, although the magnetic coupling are caused by a superexchange coupling mediated by the interfacial oxygen and nitrogen atoms in both systems, the different metal-ligand bonding configuration gives rise to different occupancy and orbital degeneracy, and hence different magnetic coupling.

## ACKNOWLEDGMENTS

This work was supported by the National Science Foundation-Material Research Science & Engineering Center (NSF-MRSEC grant #DMR-1121252). Z. L., M. Z., and F. L. acknowledge additional support from DOE-BES (Grant No. DE-FG02-04ER46148). We thank also the DOE-NERSC and the CHPC at the University of Utah for providing the computing resources.

References

[1] R. J. Curry and W. P. Gillin, Appl. Phys. Lett. **75**, 1380 (1999).

[2] A. W. Hains, Z. Q. Liang, M. A. Woodhouse, and B. A. Gregg, Chem. Rev. **110**, 6689 (2010).

[3] C. Barraud, P. Seneor, R. Mattana, S. Fusil, K. Bouzehouane, C. Deranlot, P. Graziosi, L. Hueso, I. Bergenti, V. Dediu, F. Petroff, and A. Fert, Nat. Phys. **6**, 615 (2010).

[4] S. Steil, N. Großmann, M. Laux, A. Ruffing, D. Steil, M. Wiesenmayer, S. Mathias, O. L. A. Monti, M. Cinchetti, and M. Aeschlimann, Nat. Phys. **9**, 242 (2013).

[5] Z. Y. Pang, L. Lin, F. G. Wang, S. J. Fang, Y. Dai, and S. H. Han, Appl. Phys. Lett. **99**, 153306 (2011).

[6] L. Pech, Y. A. Bankovsky, A. Kemme, and J. Lejejs, Acta Cryst. **C53**, 1043 (1997).

[7] B. Li, J. P. Zhang, X. Y. Zhang, J. M. Tian, and G. L. Huang, Inorg. Chim. Acta **366**, 241 (2011).

[8] D. Sun, C. M. Kareis, W. Jiang, G. Siegel, W. W. Shum, A. Tiwari, F. Liu, J. S. Miller, Z. V. Vardeny (submitted to Nat. Materials).

[9] H. Hu, Z. F. Wang, and F. Liu, Nanoscale Res. Lett. **9**, 690 (2014).

[10] R. Pala and F. Liu, J. Chem. Phys. **120**, 7720 (2004).

[11] A. Droghetti, S. Steil, N. Großmann, N. Haag, H. T. Zhang, M. Willis, W. P. Gillin, A. J. Drew, M. Aeschlimann, S. Sanvito, and M. Cinchetti, Phys. Rev. B **89**, 094412 (2014).

[12] Y. P. Wang and X. F. Han, Y. N. Wu and H. P. Cheng, Phys. Rev. B **85**, 144430 (2012).

[13] F. Liu, S. N. Khanna, and P. Jena, Phys. Rev. B **42**, 976 (1990).

[14] G. Kresse and D. Joubert, Phys. Rev. B **59**, 1758 (1999).

[15] J. P. Perdew and Y. Wang, Phys. Rev. B **33**, 8800 (1986).

[16] V. I. Anisimov, J. Zaanen, and O. K. Andersen, Phys. Rev. B **44**, 943 (1991).

[17] A. I. Liechtenstein, V. I. Anisimov, and J. Zaanen, Phys. Rev. B **52**, 5467 (1995).

[18] Z. Zhang and S. Satpathy, Phys. Rev. B **44**, 13319 (1991).

[19] V. I. Anisimov, I. S. Elfimov, N. Hamada, K. Terakura, Phys. Rev. B **54**, 4387 (1996).

[20] F. Ortmann, F. Bechstedt, and W. G. Schmidt, Phys. Rev. B **73**, 205101 (2006).

[21] N. Atodiresei, V. Caciuc, P. Lazic, and S. Blügel, Phys. Rev. Lett. **102**, 136809 (2009).

[22] P. Sony, P. Puschnig, D. Nabok, and C. Ambrosch-Draxl, Phys. Rev. Lett. **99**, 176401 (2007).

[23] H. J. Monkhorst and J. D. Pack, Phys. Rev. B **13**, 5188 (1976).

[24] J. B. Goodenough, *Magnetism and the Chemical Bond* (Wiley, New York, 1963).

[25] H. Wende, M. Bernien, J. Luo, C. Sorg, N. Ponpandian, J. Kurde, J. Miguel, M. Piantek, X. Xu, P. Eckhold, W. Kuch, K. Baberschke, P. M. Panchmatia, B. Sanyal, P. M. Oppeneer, and O. Eriksson, Nat. Mat. **6**, 516 (2007).

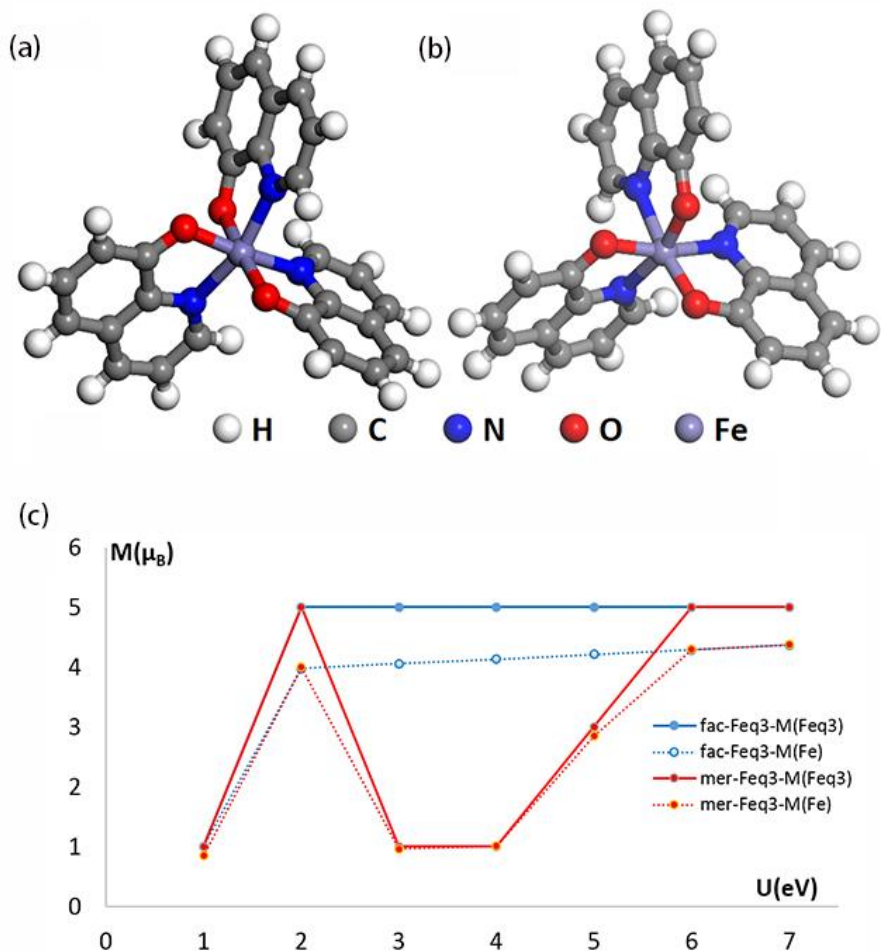

**FIG. 1.** The geometrical structure of (a) meridional Feq$_3$ and (b) facial Feq$_3$. (c) The magnetic moment as a function of the value of on-site Hubbard-U term on Fe element. The solid and dashed lines represent the magnetic moment for Feq$_3$ molecule and Fe atom respectively. The blue and red denote facial and meridional Feq$_3$ respectively.

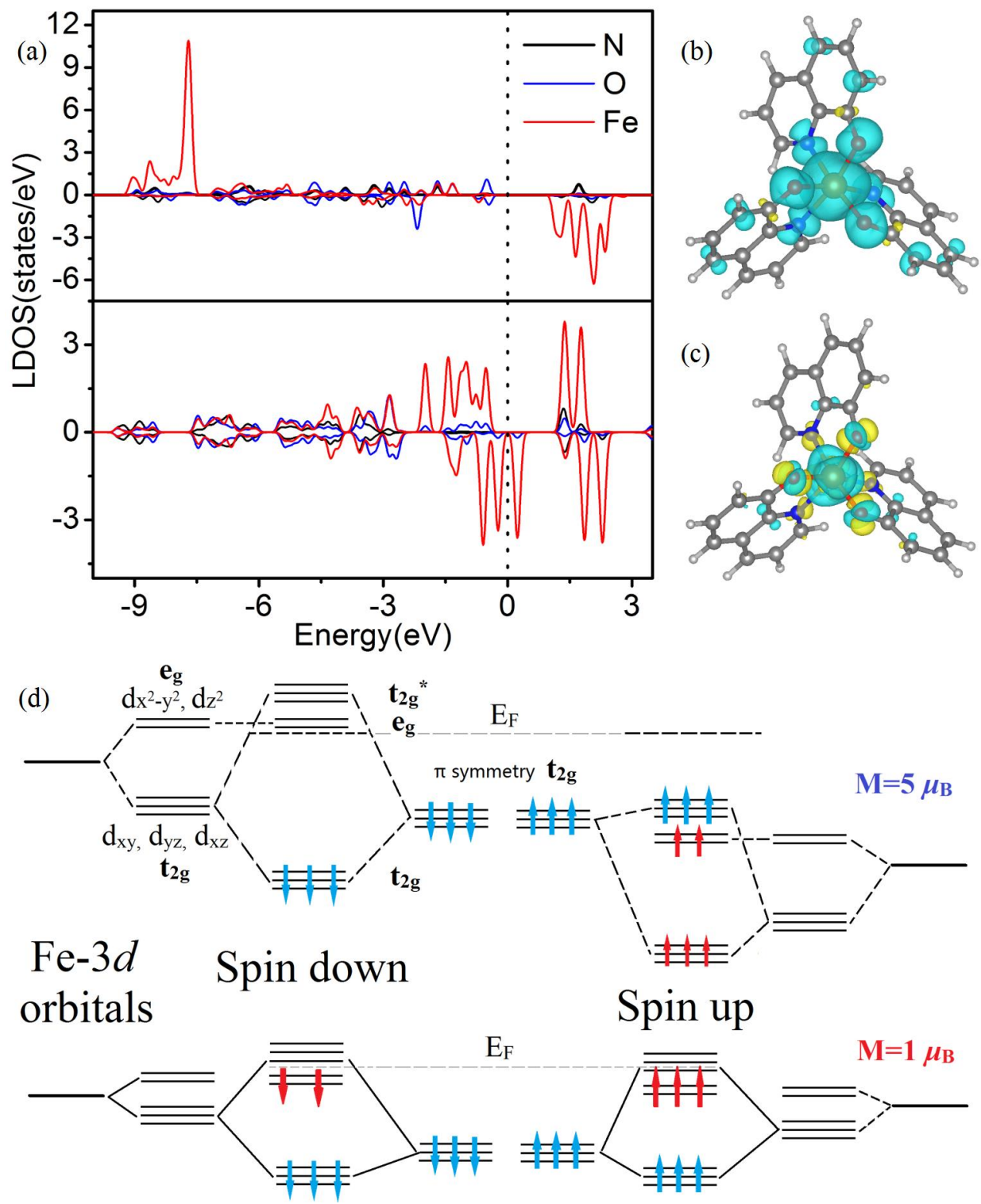

**FIG. 2.** (a) The local density of states (LDOS) of meridional Feq3 for the high spin-state (5 $\mu_B$) (upper panel) and the low spin-state (1 $\mu_B$) (lower panel). The spin distribution for (b) the high

14 | P a g e

spin-state and (c) the low spin-state; the blue and yellow denote spin-up and spin-down respectively. (d) The band structure and the distributions of electrons for the high spin-state (upper panel) and the low spin-state (lower panel); the red and blue arrows represent the electrons of center Fe atom and the neighboring ligands, respectively. The left and the right side denote spin-down and spin-up respectively.

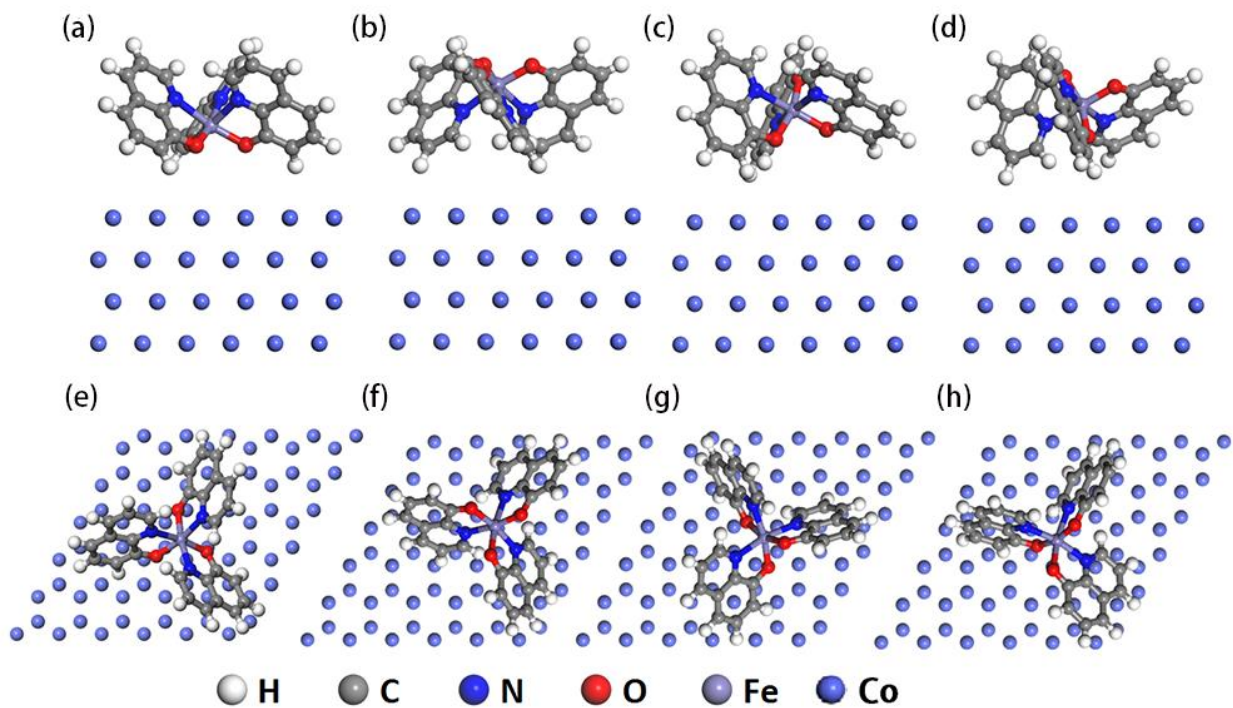

**FIG. 3.** Side (upper panel) and top (lower panel) views of adsorption configurations: (a) and (e) *fac*-O Feq$_3$/Co, (b) and (f) *fac*-N Feq$_3$/Co, (c) and (g) *mer*-O Feq$_3$/Co, (d) and (h) *mer*-N Feq$_3$/Co.

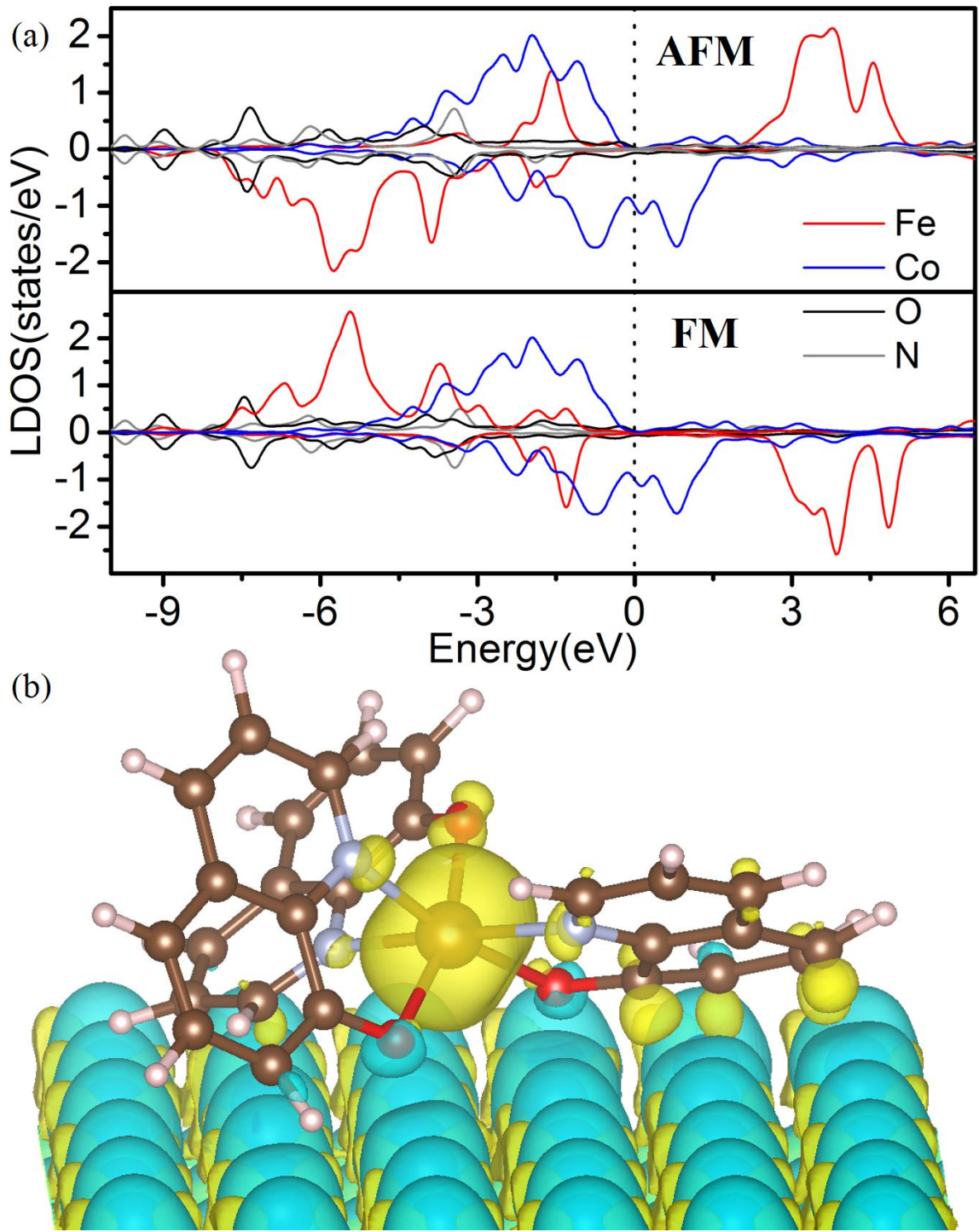

**FIG. 4.** (a) LDOS of Fe, Co, O, and N of *mer*-Feq$_3$/Co system for both AFM spin alignment (upper panel) and FM spin alignment (lower panel). (b) The spin distribution of AFM coupled Feq$_3$/Co system. Blue and yellow represent spin-up and spin-down respectively.

**TABLE I.** The bond lengths (Å) of Fe-N and Fe-O of both *fac*- and *mer*-Feq3. The experimental data are taken from Refs. 6,7.

| Bond types | *fac*-Feq$_3$ | | *mer*-Feq3 | |
|:---:|:---:|:---:|:---:|:---:|
| | Calculation | Experiment | Calculation | Experiment |
| Fe-O1 | 1.959 | 1.928 | 1.994 | 1.996 |
| Fe-O2 | 1.956 | 1.928 | 1.990 | 1.956 |
| Fe-O3 | 1.954 | 1.928 | 1.960 | 1.936 |
| Fe-N1 | 2.229 | 2.169 | 2.158 | 2.125 |
| Fe-N2 | 2.228 | 2.169 | 2.217 | 2.166 |
| Fe-N3 | 2.230 | 2.169 | 2.183 | 2.172 |

**TABLE II.** The energy differences (eV) of both facial and meridional Feq3 isomers with and without the on-site Hubbard-U term on Fe element at different fixed magnetic moments.

| Energy differences (eV) | M ($\mu_B$) | | | | |
| :---: | :---: | :---: | :---: | :---: | :---: |
| | 1 | 2 | 3 | 4 | 5 |
| *mer*-Feq$_3$ | 0 | 0.514 | 0.660 | 0.780 | 0.616 |
| *mer*-Feq$_3$ (U=3eV) | 0.091 | 0.624 | 0.358 | 0.441 | 0 |
| *fac*-Feq$_3$ | 0 | 0.489 | 0.541 | 0.597 | 0.258 |
| *fac*-Feq$_3$ (U=3eV) | 0.336 | 0.873 | 0.529 | 0.547 | 0 |

**TABLE III.** The magnetic moment of the whole Feq$_3$/Co system, the magnetic moment of Feq$_3$ molecule, and the energy differences between different structural configurations.

| Structure configuration | Energy difference (eV) | Bonding energy (eV) | Magnetic moment (total) ($\mu_B$) | Magnetic moment (Feq3) ($\mu_B$) |
|---|---|---|---|---|
| *fac*-N | 0.393 | 7.258 | 241.03 | 5 |
| *fac*-O | 0 | 7.651 | 241.19 | 5 |
| *mer*-N | 0.432 | 7.086 | 241.09 | 5 |
| *mer*-O | 0.383 | 7.135 | 241.15 | 5 |